\documentclass[sigconf]{acmart}
\AtBeginDocument{%
  }

\copyrightyear{2026}
\acmYear{2026}
\setcopyright{cc}
\setcctype{by}
\acmConference[CIKM '26]{Proceedings of the 35th ACM International Conference on Information and Knowledge Management}{November 07--11, 2026}{Rome, Italy}
\acmBooktitle{Proceedings of the 35th ACM International Conference on Information and Knowledge Management (CIKM '26), November 07--11, 2026, Rome, Italy}
\acmDOI{10.1145/3799682.3841124}
\acmISBN{979-8-4007-2539-5/2026/11}

\usepackage{multirow}
\usepackage{algorithm}
\usepackage{algpseudocode}
\newcommand{\rup}[1]{\rlap{$_{\textcolor{red}{\uparrow #1\%}}$}}
\begin{document}

\title{Faithful Evaluation of Semantic-ID Tokenizers for Generative Recommendation}

\author{Qian Zhang}
\authornote{Corresponding author.}
\orcid{0009-0009-1117-0236}
\email{zhaqi075@student.otago.ac.nz}
\affiliation{%
  \department{School of Computing}
  \institution{University of Otago}
  \city{Dunedin}
  \country{New Zealand}
}

\author{Lech Szymanski}
\orcid{0000-0002-5192-0304}
\email{lech.szymanski@otago.ac.nz}
\affiliation{%
  \department{School of Computing}
  \institution{University of Otago}
  \city{Dunedin}
  \country{New Zealand}
}

\author{Jeremiah D. Deng}
\orcid{0000-0003-3727-4403}
\email{jeremiah.deng@otago.ac.nz}
\affiliation{%
  \department{School of Computing}
  \institution{University of Otago}
  \city{Dunedin}
  \country{New Zealand}
}

\author{Haibo Zhang}
\orcid{0000-0002-3752-0806}
\email{haibo.zhang@unsw.edu.au}
\affiliation{%
  \department{School of Computer Science and Engineering}
  \institution{University of New South Wales}
  \city{Sydney}
  \country{Australia}
}

\renewcommand{\shortauthors}{Qian Zhang, Lech Szymanski, Jeremiah D. Deng, and Haibo Zhang}

\begin{abstract}
  Generative recommendation based on Semantic IDs (SIDs) represents each item as a discrete sequence of SIDs and is conventionally evaluated by matching the generated SID sequence against the target item's SID sequence. This evaluation is faithful only when each SID sequence uniquely identifies one item. In practice, SID collisions violate this condition: across the evaluated SID tokenizers and datasets, collision rates reach 30.52\%, and SID-level Hit@10 is inflated by up to 103.36\% relative to item-level Hit@10. To address this evaluation gap, we introduce Collision-Corrected Evaluation (CCE), which defines collision-aware item-level metrics (ItemHit@$K$, ItemNDCG@$K$) computed from generated SID sequences, and Zero-Collision Reassignment (ZCR), which constructs zero-collision SID assignments for existing tokenizers via minimum-cost reassignment. Applying these methods to four datasets and five representative SID tokenizers, we find that metric inflation scales with collision rate and is large enough to flip pairwise tokenizer comparisons under item-level re-evaluation. This finding calls into question the reliability of SID-level rankings reported in prior work and indicates that faithful tokenizer evaluation requires item-level correction or zero-collision SID assignments.
\end{abstract}




\maketitle

\section{Introduction}
\label{sec:intro}

Generative recommendation has emerged as a paradigm that represents items as Semantic ID (SID) sequences and trains an autoregressive model to generate the SID sequence of the next item from a user's interaction history \cite{TIGER2023,OneRec2025,LLM4RecSurvey2025,TallRec2023}. In this paradigm, item tokenization plays a central role by mapping items to discrete code sequences that serve as the generator's training targets \cite{LETTER2024,EAGER2024}. Consequently, recent studies have increasingly focused on quantization-based tokenizers and recommendation-aware tokenizer objectives to improve the quality of SID representations \cite{ReSID2026,QuaSID2026,ETEGRec2025,FACE2025,ScalingSID2025}. However, the evaluation protocols used to compare these tokenizers implicitly assume that each generated SID sequence uniquely identifies one item.

\begin{figure}[!t]
  \centering
  \includegraphics[width=\columnwidth]{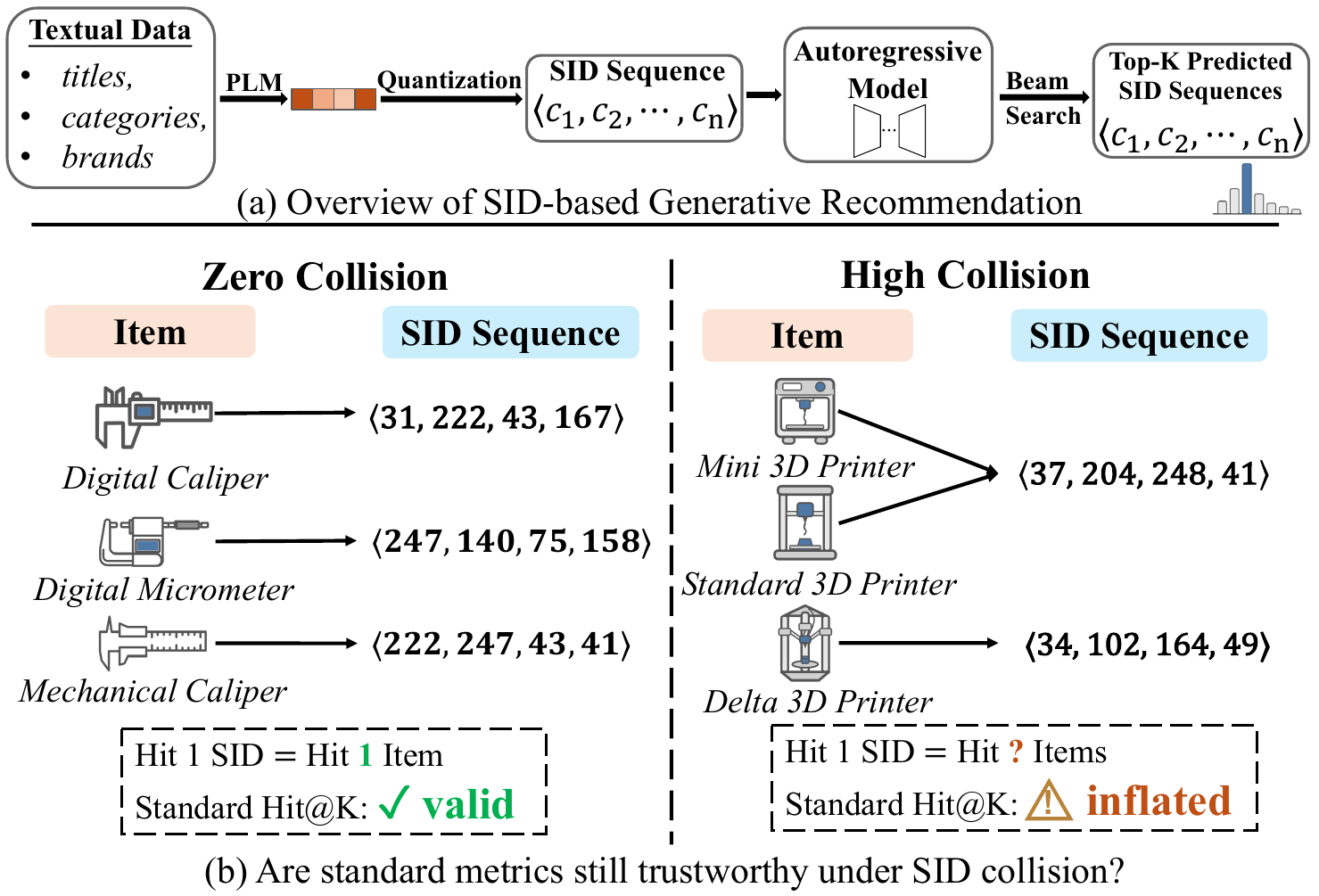}
  \caption{(a) Overview of SID-based generative recommendation. (b) Under zero collision, each SID sequence identifies one item; under SID collision, the same SID sequence maps to multiple items, inflating estimates of item-level performance.}
  \Description{Two-panel diagram illustrating Semantic ID generation and evaluation ambiguity caused by SID collisions. Panel (a) shows item metadata encoded by a pretrained language model, quantized into Semantic ID sequences, and used by an autoregressive model with beam search to generate top-K predictions. Panel (b) contrasts zero collision, where each item maps to a distinct Semantic ID, with SID collision, where multiple 3D printers share the same sequence. Under zero collision, a matched Semantic ID uniquely identifies an item; under collision, it can correspond to multiple items, causing standard Hit at K to overestimate item-level performance.}
  \label{fig:sid_collision}
\end{figure}

Common evaluation protocols in SID-based generative recommendation compute top-$K$ metrics by checking whether the SID sequence of the target item appears in the ranked list of generated SID sequences \cite{LETTER2024,QuaSID2026,ETEGRec2025}, as illustrated in Figure~\ref{fig:sid_collision}(a). However, recommendation accuracy is fundamentally item-level: a generated SID sequence should count as correct only if it uniquely maps to the target item. \emph{SID collision}, where multiple items share the same SID sequence, breaks this unique mapping \cite{QuaSID2026}. In this case, the generated sequence identifies only a collision group rather than the target item itself, causing SID-level metrics to overestimate item-level performance and bias tokenizer comparisons.

Figure~\ref{fig:sid_collision}(b) contrasts zero collision with SID collision: the generated SID sequence maps to a single item under zero collision, but to a group of items under SID collision. For example, when two similar 3D printers (\emph{Mini 3D Printer} and \emph{Standard 3D Printer}) share one SID sequence, generating that SID cannot distinguish which is the target item. Thus, what appears to be a correct SID prediction may only be a group-level match, motivating faithful item-level evaluation of generated SID lists. Otherwise, tokenizers with higher collision rates can be favored by inflated SID-level metrics.
Table~\ref{tab:collision_stats} further shows that such collisions are not rare and occur across datasets and tokenizers. Collision rates range from 8.52\% (Beauty, RQ-VAE) to 30.52\% (Scientific, RK-Means), with RK-Means consistently producing higher rates than RQ-VAE on the same dataset. Because these collisions occur in practice, matching the target SID sequence does not always mean identifying the target item. Therefore, faithful evaluation should assign full item-level credit only when a generated SID sequence identifies the target item, and adjust the credit when SID collision occurs.

\begin{table}[t]
\centering
\caption{SID collision in two representative SID tokenizers across four datasets. Coll.\% denotes the fraction of items whose SID sequence is shared with at least one other item, and $G_{\max}$ denotes the size of the largest collision group.}
\label{tab:collision_stats}
\begin{tabular}{llrr}
\toprule
Dataset & Tokenizer & Coll.\% & $G_{\max}$ \\
\midrule
Scientific & RK-Means & 30.52 & 26 \\
Scientific & RQ-VAE    & 20.42 & 22 \\
Cell       & RK-Means & 28.52 & 56 \\
Cell       & RQ-VAE    & 14.04 & 34 \\
Beauty     & RK-Means & 21.58 & 18 \\
Beauty     & RQ-VAE    &  8.52 & 12 \\
Yelp       & RK-Means & 15.73 & 65 \\
Yelp       & RQ-VAE    &  9.05 & 65 \\
\bottomrule
\end{tabular}
\end{table}

For faithful item-level evaluation under SID collision, we introduce Collision-Corrected Evaluation (CCE) and use ItemHit@$K$ and ItemNDCG@$K$ as the main metrics, distinguishing them from SID-level Hit@$K$ and NDCG@$K$. When a matched SID sequence is shared across multiple items, these metrics divide the item-level credit by the collision-group size because the SID sequence no longer uniquely identifies a single item. We also report conventional Hit@$K$ and NDCG@$K$ as SID-level reference metrics for comparability with previous work.

CCE provides faithful item-level scores even when SID collisions remain. However, CCE still evaluates each tokenizer on its native SID outputs, whose collision rates range from 8.52\% to 30.52\% in Table~\ref{tab:collision_stats}. Thus, tokenizers are still compared under different degrees of SID collision. To compare tokenizers without SID collisions affecting the comparison, we need a zero-collision SID assignment that stays close to each tokenizer's original assignment; otherwise, performance differences may be caused by excessive changes to the SID assignment rather than by removing collisions. To this end, we introduce Zero-Collision Reassignment (ZCR), a post-tokenizer method that constructs a zero-collision SID assignment with minimum reassignment cost.

In summary, our contributions are as follows:
    
\begin{itemize}
    \item We identify \emph{collision-induced metric inflation}, where SID-level metrics overestimate item-level recommendation performance under SID collision.
    
    \item We introduce CCE, using ItemHit@$K$ and ItemNDCG@$K$ to measure faithful item-level performance from the generated SID sequences.

    \item We introduce ZCR, a post-tokenizer method that constructs zero-collision SID assignments with minimum reassignment cost.

    \item We conduct experiments across four datasets and five representative SID tokenizers, demonstrating that higher collision rates result in larger metric inflation and that this inflation is large enough to flip pairwise tokenizer comparisons under item-level re-evaluation. For reproducibility, our implementation and the pre-built SID indexes for all tokenizers and datasets are available at \url{https://github.com/Nishikata97/CollisionGenRec}.
\end{itemize}

\section{Task Formulation}
\label{sec:task_formulation}

\subsection{SIDs-based Generative Recommendation}
Let $\mathcal{U}$ and $\mathcal{I}$ denote the user set and item set, respectively, where $|\mathcal{I}| = N$ is the number of items. Each user has a chronological interaction history $(i_1, i_2, \ldots, i_t)$, where $i_t \in \mathcal{I}$ denotes the item interacted with at time step $t$, and the task is to predict the next item $i_{t+1} \in \mathcal{I}$ (target item).

SID-based generative recommendation comprises two core components: item tokenization and SID-based generation. 

\emph{(i) Item Tokenization.} Each item $i \in \mathcal{I}$ is mapped by a tokenizer to a fixed length-$L$ discrete code sequence (i.e., SID sequence):
\begin{equation}
\label{eq:sid}
\mathbf{s}_i = [s_i^{(1)}, \ldots, s_i^{(L)}] \in \{1, \ldots, V\}^L, \quad i \in \mathcal{I},
\end{equation}
where $V$ is the codebook size and $L$ is the length of SID sequence.  At level $l$, let $\mathcal{E}^{(l)}=\{\mathbf{e}_1^{(l)},\ldots,\mathbf{e}_V^{(l)}\}$ denote the tokenizer's codebook, so $s_i^{(l)}$ indexes the codebook vector $\mathbf{e}_{s_i^{(l)}}^{(l)}$. Once trained, the tokenizer's mapping is frozen, forming an item-to-SID lookup table from $\mathcal{I}$ to the SID space ${1, \ldots, V}^L$. Since $|\mathcal{I}| = N$ items are mapped into an SID space of size $V^L$, \emph{collisions are possible when $N$ approaches $V^L$}.

\emph{(ii) SID-based Generation.} After offline tokenization, user interaction $(i_1, i_2, \ldots, i_t)$ is represented in SID form as $(\mathbf{s}_{i_1}, \ldots, \mathbf{s}_{i_t})$, and the generator predicts the target SID sequence $\mathbf{s}_{i_{t+1}}$ conditioned on the sequence of historical SIDs. At inference time, beam search returns ranked SID sequences rather than uniquely identified items.

\subsection{SID Collision}
Vector quantization~\cite{VQVAE2017,PQ2011} was originally developed for representation compression, allowing different representations to share the same code. However, SID-based generative recommendation repurposes quantization as an item identifier. Since the recommendation target is fundamentally an item, faithful evaluation requires the tokenizer to establish a one-to-one mapping between items and SID sequences. Traditional vector quantization does not impose this requirement, leading to multiple items sharing the same SID sequence (\emph{SID collision}). Formally, a SID collision occurs when there exist two distinct items $i \neq j$ such that $\mathbf{s}_i = \mathbf{s}_j$. For any SID sequence $\mathbf{s}$, we define its collision group as
\begin{equation}
  \mathcal{C}(\mathbf{s}) = \{i \in \mathcal{I} : \mathbf{s}_i = \mathbf{s}\}.
\end{equation}
When $|\mathcal{C}(\mathbf{s})| > 1$, a generated SID identifies a collision group rather than a unique item. We define the collision rate as the percentage of items whose SID sequence is shared with at least one other item:
\begin{equation}
  \mathrm{Coll.}\% =
  \frac{
    |\{ i \in \mathcal{I} : |\mathcal{C}(\mathbf{s}_i)| > 1 \}|
  }{N}
  \times 100,
\end{equation}
where $|\cdot|$ denotes the number of elements in a set.

\textbf{Collision-aware Evaluation.} Under SID collision, exact SID matching no longer provides faithful item-level evaluation, because a matched SID may correspond to a collision group rather than a unique item. Section~\ref{sec:cce} formalizes the corresponding collision-aware evaluation metrics.

\section{Methodology}
To enable faithful tokenizer comparison under SID collisions, we propose collision-aware evaluation metrics and a zero-collision SID reassignment method that jointly close the evaluation gap. \textbf{Collision-Corrected Evaluation} (CCE, Section~\ref{sec:cce}) addresses the measurement bias induced by SID collisions by introducing ItemHit@$K$ and ItemNDCG@$K$ as corrected item-level metrics computed directly from saved beam outputs without retraining or modifying the tokenizer. However, even with corrected metrics, the autoregressive model is still trained on the tokenizer's native SID assignment containing collisions (Table~\ref{tab:collision_stats}). Cross-tokenizer comparison remains confounded by varying collision rates in the training targets. \textbf{Zero-Collision Reassignment} (ZCR, Section~\ref{sec:zcr}) addresses collisions by constructing zero-collision SID assignments at minimum reassignment cost, producing a uniform setting for cross-tokenizer comparison. Together, CCE corrects measurement under any degree of SID collision, and ZCR provides a zero-collision SID assignment for generator training, jointly enabling faithful tokenizer comparison.

\subsection{Collision-Corrected Evaluation}
\label{sec:cce}

Standard generative recommendation evaluation reports SID-level Hit@$K$ and NDCG@$K$ on the ranked list of predicted SID sequences from beam search~\cite{LETTER2024,QuaSID2026,ETEGRec2025}. As shown in Figure~\ref{fig:sid_collision}, this SID-level protocol inflates item-level recommendation performance under SID collision~\cite{LCRec2024,QuaSID2026}, because a matched SID sequence may identify a collision group rather than the target item. To align the evaluation with the item-level recommendation objective, CCE expands each predicted SID sequence to all items in the corresponding collision group and assigns uniform fractional credit, rather than treating the SID match as a full item-level hit.

\begin{figure}[t]
  \centering
  \includegraphics[width=0.9\columnwidth]{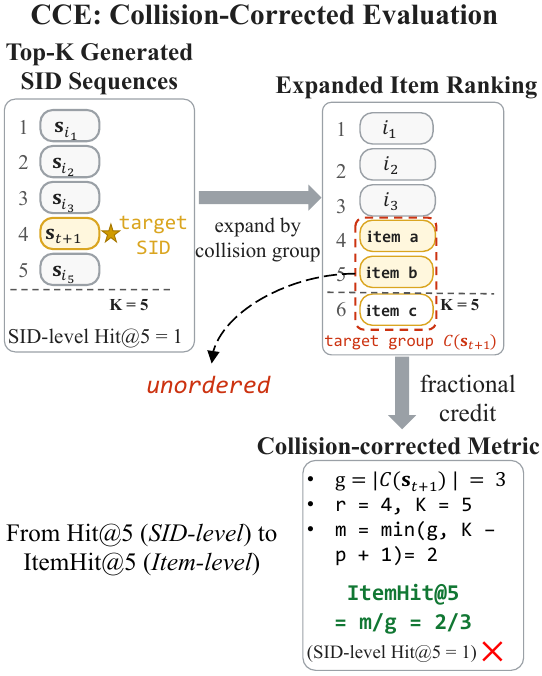}
  \caption{Collision-corrected evaluation. The expanded item ranking determines how a SID-level hit is credited under item-level top-$K$ evaluation.}
  \Description{Flow diagram of a collision-corrected top-five evaluation example. Five generated Semantic ID sequences are ranked on the left, with the target sequence at rank four. Expanding the ranking to items places the three unordered items sharing the target Semantic ID at positions four through six. Because the top-five cutoff includes only two of the three items, the prediction receives item-level credit of two thirds rather than full Semantic ID-level credit.}
  \label{fig:cce}
\end{figure}

For a user with history $(i_1,\ldots,i_t)$, the task is to predict the target item $i_{t+1}$, whose SID sequence we denote $\mathbf{s}_{i_{t+1}}$. Beam search returns a top-$K$ ranked list $B=(\mathbf{s}^{(1)},\ldots,\mathbf{s}^{(K)})$ of candidate SID sequences. We refer to the collision group of the target SID as the \emph{target group}, with size $g = |\mathcal{C}(\mathbf{s}_{i_{t+1}})|$. Let $r$ denote the rank at which the target SID sequence $\mathbf{s}_{i_{t+1}}$ first appears in $B$. If it does not appear in $B$, set $m=0$. Otherwise, we form a single expanded item ranking by concatenating the collision groups in beam order, $\mathcal{C}(\mathbf{s}^{(1)}), \mathcal{C}(\mathbf{s}^{(2)}), \ldots$. All items in the same collision group are unordered (the dashed link in Figure~\ref{fig:cce}), since SID generation alone cannot distinguish them. The target group then occupies the expanded positions
\begin{equation}
  p,\; p+1,\; \ldots,\; p+g-1,
  \quad
  p = 1 + \sum_{q<r} |\mathcal{C}(\mathbf{s}^{(q)})|,
\end{equation}
where $p$ denotes the starting position of the target group in the expanded item ranking.

The number of items in the target group that fall before the top-$K$ cutoff is therefore
\begin{equation}
  m = \min(g,\max\{0,K-p+1\}).
\end{equation}
The formula also covers $p>K$: the collision groups before the target group may contain more than $K-1$ items in total, pushing the target group beyond the top-$K$ cutoff so that $m=0$.
 

We then define the collision-corrected metrics as follows:
\begin{equation}
  \text{ItemHit@}K = \frac{m}{g},
  \label{eq:itemhit}
\end{equation}
Figure~\ref{fig:cce} illustrates this with $K=5$: the target SID sequence appears at rank $r=4$, its target group has $g=3$ items, and only $m=2$ of them fall in the top-$5$ expanded positions, yielding $\text{ItemHit@}5 = 2/3$ instead of the SID-level $\text{Hit@}5 = 1$. ItemHit@$K = m/g$ requires no further normalization because $m \le g$ ensures the value lies in $[0, 1]$, equivalent to the mean $\text{Hit}$ over the $g$ possible ranks of the target item within its group. Note that in next-item prediction each test sample has a single target item, so Hit@$K$ is equivalent to Recall@$K$~\cite{BERT4Rec2019,SampledMetrics2020}, and ItemHit@$K$ can therefore equivalently be read as a collision-corrected Recall@$K$.

To account for position in the top-$K$, ItemNDCG@$K$ sums the standard log-discount over the target items within the top-$K$, averaged by $g$:
\begin{equation}
  \text{ItemNDCG@}K = \frac{1}{g} \sum_{e=1}^{m} \frac{1}{\log_2(p + e)}.
  \label{eq:itemndcg}
\end{equation}
In $\text{ItemNDCG}$, the $m$ items at expanded positions $p, p+1, \ldots, p+m-1$ each contribute the standard NDCG log-discount: the $e$-th item at position $p+e-1$ contributes $1/\log_2((p+e-1)+1) = 1/\log_2(p+e)$. No IDCG normalization is needed because each test sample has a single target item, so the ideal ranking places the target item at position $1$ and $\text{IDCG} = 1/\log_2 2 = 1$. For the case in Figure~\ref{fig:cce}, this yields $\text{ItemNDCG@}5 = (1/\log_2 5 + 1/\log_2 6)/3 \approx 0.273$, compared with the SID-level $\text{NDCG@}5 = 1/\log_2 5 \approx 0.431$. When $g = 1$, the target SID identifies a unique item, $m \in \{0, 1\}$, and both ItemHit@$K$ and ItemNDCG@$K$ reduce to the standard Hit@$K$ and NDCG@$K$. When $g > 1$, SID-level metrics can exceed their item-level counterparts. Table~\ref{tab:collision_inflation} quantifies this inflation empirically.

\subsection{Zero-Collision Reassignment}
\label{sec:zcr}

To preserve the hierarchical structure of the native tokenizer while resolving collisions, ZCR keeps the first $L-1$ codewords of each item unchanged and reassigns only the last-level codeword. This design is inspired by TIGER, which appends an extra disambiguating code to the SID tail to resolve collisions~\cite{TIGER2023}, although the appended code does not carry semantic information. Last-level reassignment is sufficient when the last-level codebook has enough codes to assign distinct last-level codes to all items in each prefix group (Eq.~\ref{eq:codebook_capacity}). This last-level codebook capacity condition is satisfied across all four datasets and four tokenizers in our experiments with $V=256$ (Appendix~\ref{sec:capacity-appendix}). Among assignments that produce unique SIDs under this last-level-only constraint, ZCR first minimizes the number of changed last-level codewords and then minimizes the total reassignment cost (Eq.~\ref{eq:cr_cost}).

Given SID assignments produced by a tokenizer, ZCR groups items sharing the same length-$(L-1)$ prefix $\mathbf{p} = \mathbf{s}_i^{1:L-1}$ into a prefix group $\mathcal{P}(\mathbf{p})$. Since the first $L{-}1$ codewords are preserved, collisions only occur within a prefix group, and reassigning the last-level code $s_i^{(L)}$ decomposes into independent per-group subproblems. For each prefix group $\mathcal{P}(\mathbf{p})$, the minimum number of last-level changes needed for all its items to have distinct last-level codes is
\begin{equation}
  \label{eq:rho_def}
  \rho_{\mathbf{p}} = |\mathcal{P}(\mathbf{p})| - |\{s_i^{(L)} : i \in \mathcal{P}(\mathbf{p})\}|,
\end{equation}
where $|\mathcal{P}(\mathbf{p})|$ is the number of items in the prefix group and $|\{s_i^{(L)} : i \in \mathcal{P}(\mathbf{p})\}|$ is the number of distinct last-level codes among them. The total number of last-level changes thus equals $\sum_{\mathbf{p}} \rho_{\mathbf{p}}$ regardless of the order in which groups are processed.

ZCR processes only the prefix groups where reassignment is both needed and possible. We denote this set as
\begin{equation}
  \label{eq:zcr_act_set}
  \mathcal{P}_{\text{coll}} = \{\, \mathbf{p} \,\mid\, \rho_{\mathbf{p}} > 0,\; |\mathcal{P}(\mathbf{p})| \le V \,\},
\end{equation}
where $\rho_{\mathbf{p}} > 0$ requires the prefix group to contain at least one collision, and $|\mathcal{P}(\mathbf{p})| \le V$ is the exact condition for a distinct last-level assignment: with more than $V$ items, two of them must share a code since only $V$ codes are available, while with at most $V$ items, each can receive its own code. ZCR processes the prefix group $\mathcal{P}(\mathbf{p})$ for each $\mathbf{p} \in \mathcal{P}_{\text{coll}}$ and leaves other groups unchanged: those that are already collision-free ($\rho_{\mathbf{p}} = 0$) or too large for distinct last-level codes ($|\mathcal{P}(\mathbf{p})| > V$). The latter case does not arise in our experiments (Appendix~\ref{sec:capacity-appendix}), and the last-level scope is chosen because ZCR guarantees the global minimum reassignment cost only in this scope (Section~\ref{sec:optimality}).

To formalize this minimum-cost objective, ZCR uses the squared distance between the last-level residual $\mathbf{r}_i^{(L-1)}$ of item $i$ and each last-level candidate codebook vector $\mathbf{e}_c^{(L)}$ as the per-code cost: $D[i, c] = \|\mathbf{r}_i^{(L-1)} - \mathbf{e}_c^{(L)}\|^2$, where both vectors come from the tokenizer's quantization rather than the learned embeddings of generator. For each $\mathbf{p} \in \mathcal{P}_{\text{coll}}$ (Eq.~\ref{eq:zcr_act_set}), ZCR finds reassigned last-level codes $\hat{s}_i^{(L)} \in \{1,\ldots,V\}$ for items $i$ in the prefix group $\mathcal{P}(\mathbf{p})$ by minimizing the total reassignment cost:
\begin{equation}
\label{eq:cr_cost}
\begin{aligned}
\min_{\{\hat{s}_i^{(L)}\}}
& \sum_{i \in \mathcal{P}(\mathbf{p})}
\bigl(D[i,\hat{s}_i^{(L)}] - D[i,s_i^{(L)}]\bigr) \\
\text{s.t.}\
& |\{i : \hat{s}_i^{(L)} \ne s_i^{(L)}\}| = \rho_{\mathbf{p}},\
\hat{s}_i^{(L)} \ne \hat{s}_j^{(L)} \text{ for } i \ne j.
\end{aligned}
\end{equation}
Eq.~\ref{eq:cr_cost} is a constrained min-cost bipartite assignment between $|\mathcal{P}(\mathbf{p})|$ items and $V$ last-level codes, with $\rho_\mathbf{p}$ items changing their last-level code and all items in the prefix group receiving distinct codes. The Algorithm~\ref{alg:zcr} solves this using the Hungarian algorithm~\cite{Hungarian1955} in polynomial time, and Section~\ref{sec:optimality} establishes its optimality scope.

\paragraph{Last-level codebook capacity.}
Formally, this sufficiency condition requires the last-level codebook size $V$ to upper-bound every prefix-group size:
\begin{equation}
    \label{eq:codebook_capacity}
    \max_{\mathbf{p}} |\mathcal{P}(\mathbf{p})| \le V.
\end{equation}
Under this condition, items in different prefix groups already differ in their fixed prefixes, and items within each prefix group can be assigned distinct last-level codes. Therefore, the reassigned SID assignment is globally collision-free. This condition holds across all tokenizers and datasets in our experiments with $V=256$ (Appendix~\ref{sec:capacity-appendix}), and Section~\ref{sec:optimality} justifies restricting reassignment to the last level over multi-level assignment.

\begin{algorithm}[t]
  \caption{Zero-Collision Reassignment}
  \label{alg:zcr}
  \begin{algorithmic}[1]
  \Require Native SID assignment $\{\mathbf{s}_i\}_{i \in \mathcal{I}}$; last-level residuals $\{\mathbf{r}_i^{(L-1)}\}_{i \in\mathcal{I}}$; last-level codebook $\{\mathbf{e}_c^{(L)}\}_{c=1}^V$
  \Ensure Reassigned SIDs $\{\hat{\mathbf{s}}_i\}_{i \in \mathcal{I}}$; collision-free if Eq.~\ref{eq:codebook_capacity} is satisfied
  
  \State Initialize $\hat{\mathbf{s}}_i \gets \mathbf{s}_i$ for all $i \in \mathcal{I}$
  \State Compute $D[i, c] \gets \|\mathbf{r}_i^{(L-1)} - \mathbf{e}_c^{(L)}\|^2$ for all $i \in \mathcal{I}$, $c \in \{1,\ldots,V\}$
  \State Partition items by native $(L{-}1)$-prefix: $\mathcal{P}(\mathbf{p}) \gets \{i : \mathbf{s}_i^{1:L-1} = \mathbf{p}\}$
  \For{each prefix group $\mathcal{P}(\mathbf{p})$}
    \State $\rho_{\mathbf{p}} \gets |\mathcal{P}(\mathbf{p})| - |\{s_i^{(L)} : i \in \mathcal{P}(\mathbf{p})\}|$
    \If{$\rho_{\mathbf{p}} = 0$ \textbf{or} $|\mathcal{P}(\mathbf{p})| > V$}
      \State \textbf{continue}
    \EndIf
    \State Obtain $\{\hat{s}_i^{(L)}\}_{i \in \mathcal{P}(\mathbf{p})}$ by solving Eq.~\ref{eq:cr_cost} on $\mathcal{P}(\mathbf{p})$
  \EndFor
  \State \Return $\{\hat{\mathbf{s}}_i\}_{i \in \mathcal{I}}$
  \end{algorithmic}
\end{algorithm}

\subsubsection{Algorithm and Complexity}
\label{sec:algorithm}
Algorithm~\ref{alg:zcr} implements ZCR through independent per-group optimization, enabled by the prefix-group decomposition. In each group, we solve Eq.~\ref{eq:cr_cost} with the Hungarian algorithm after adding a large $M$ to each cost $D[i,c]$ with $c \ne s_i^{(L)}$, thereby enforcing its count constraint, in $O(|\mathcal{P}(\mathbf{p})|^2 V)$; the total complexity $O(NVd + \sum_{\mathbf{p}} |\mathcal{P}(\mathbf{p})|^2 V)$, where $N = |\mathcal{I}|$ is the number of items and $d$ is the residual dimension, is dominated by the distance matrix computation, computed in batch with the FAISS library~\cite{FAISS2021}. In our experiments ($N \le 5 \cdot 10^4$, $V = 256$), the runtime of ZCR is negligible relative to tokenizer training.

\subsubsection{Optimality and Scope of Last-level Reassignment}
\label{sec:optimality}

Eq.~\ref{eq:cr_cost} is a constrained min-cost optimization, solvable in each fixed prefix group as a bipartite assignment between $|\mathcal{P}(\mathbf{p})|$ items and $V$ last-level codes, in which exactly $\rho_{\mathbf{p}}$ items change their last-level code and all items in the group receive distinct last-level codes. Because each prefix group's constraints involve only its own items and the total cost is additive across groups, per-group optimal solutions combine into the global minimum over assignments that preserve the first $L{-}1$ codewords and modify only the last level.

A natural extension would allow earlier codewords, such as $s_i^{(L-1)}$ or $s_i^{(L-2)}$, to change together with $s_i^{(L)}$. Such multi-level reassignment can theoretically expand the search space, but it loses two properties that make ZCR both tractable and aligned with the faithful-evaluation goal of this work.

\begin{itemize}
\item First, the per-group decomposition fails. ZCR fixes the first $L{-}1$ codewords of every item, so the prefix-group partition is fixed in advance, each group contains at most $V$ items (Eq.~\ref{eq:codebook_capacity}), and Eq.~\ref{eq:cr_cost} is solved one group at a time. Allowing the last $n$ levels to change requires coarsening the partition to the $(L{-}n)$-prefix level, where the per-item search space grows exponentially in $n$ as $V^n$. The per-group decomposition of ZCR is intrinsic to varying only the last level ($n=1$).

\item Second, modifying earlier codewords changes the residuals passed to subsequent levels. Residual quantization is sequential: the tokenizer constructs $\mathbf{r}_i^{(\ell)} = \mathbf{r}_i^{(\ell-1)} - \mathbf{e}^{(\ell)}_{s_i^{(\ell)}}$ during training. Therefore, changing an earlier assignment such as $s_i^{(L-1)}$ would also change $\mathbf{r}_i^{(L-1)}$, which is the residual used to score last-level assignments in Eq.~\ref{eq:cr_cost}. In that case, the reassignment cost would no longer be computed in the tokenizer's original residual space. ZCR avoids this issue by modifying only the last level, preserving the native prefix structure and evaluating alternative last-level codewords based on the residuals produced by the original tokenizer.
\end{itemize}

The capacity condition in Eq.~\ref{eq:codebook_capacity} is generously satisfied in our experiments: all $\max_\mathbf{p}|\mathcal{P}(\mathbf{p})|$ lie well below $V=256$ (Appendix~\ref{sec:capacity-appendix}). This slack reflects the overcapacity of the SID space: the configuration $L=4, V=256$ adopted by previous SID-based generative recommendation work~\cite{TIGER2023,LETTER2024,MQL4GRec2025} and used here yields $V^L \approx 4.3\times10^9$ distinct sequences, roughly five orders of magnitude beyond our largest dataset (Table~\ref{tab:datasets}). Therefore, the collision rates of $8.52\%$--$30.52\%$ in Table~\ref{tab:collision_stats} depend on both the design of tokenizers and the text quality of datasets: RK-Means consistently produces higher rates than RQ-VAE on the same dataset, while the largest Yelp collision group of size $65$ appears identically under both tokenizers. The capacity condition fails only when near-identical items exceed $V$ in count, which does not occur for the four datasets. ZCR produces zero-collision SID assignments for all four collision-prone tokenizers across all four datasets.

\section{Experiments}

We conduct extensive experiments to evaluate CCE and ZCR for faithful comparison of SID tokenizers. Specifically, we focus on the following research questions:

\begin{itemize}
    \item [\textbf{RQ1:}] To what extent do SID collisions inflate conventional metrics (\emph{SID-level}) compared to collision-corrected metrics (\emph{item-level})?

    \item [\textbf{RQ2:}] Does applying CCE change the relative ranking of SID tokenizers compared to conventional SID-level metrics?

    \item [\textbf{RQ3:}] Can ZCR construct zero-collision SID assignments with low reassignment cost, and how does this affect downstream item-level performance?
\end{itemize}

\subsection{Experimental Settings}
\subsubsection{Datasets}
We conduct experiments on four widely used public recommendation datasets: \textit{Scientific} (\textit{Industrial and Scientific}), \textit{Cell} (\textit{Cell Phones and Accessories}), and \textit{Beauty} are review datasets from the Amazon Review collection~\cite{Amazon2019}, and \textit{Yelp} is a local-business review dataset\footnote{\href{https://business.yelp.com/data/resources/open-dataset/}{https://business.yelp.com/data/resources/open-dataset/}}. We select these four datasets to examine collision inflation and tokenizer quality across different recommendation domains and sparsity levels. Following prior SID-based generative recommendation protocols~\cite{TIGER2023,LETTER2024,LCRec2024,GRID2025}, we apply 5-core filtering to the Amazon and Yelp datasets. We split each user's chronological sequence by leave-one-out, using the last interaction for testing, the second-to-last interaction for validation, and the remaining interactions for training. We construct textual item features from available metadata fields: \textit{title}, \textit{description}, \textit{brand}, and \textit{categories} for Amazon; \textit{business name} and \textit{categories} for Yelp. Dataset statistics are reported in Table~\ref{tab:datasets}.

\begin{table}[t]
  \centering
  \caption{Statistics of the processed datasets.}
  \label{tab:datasets}
  \begin{tabular*}{\columnwidth}{@{\extracolsep{\fill}}lrrrrr@{}}
    \toprule
    \textbf{Dataset} & \textbf{\#Users} & \textbf{\#Items} & \textbf{\#Inter.} & \textbf{Avg.\,Len.} & \textbf{Sparsity} \\
    \hline
    Scientific &   8{,}317 &  4{,}344 &      58{,}492 &  7.03 & 99.84\% \\
    Cell       & 154{,}778 & 47{,}593 &   1{,}109{,}257 &  7.17 & 99.98\% \\
    Beauty     &  22{,}363 & 12{,}101 &     198{,}502 &  8.88 & 99.93\% \\
    Yelp       &  30{,}431 & 20{,}033 &     316{,}354 & 10.40 & 99.95\% \\
    \bottomrule
  \end{tabular*}
\end{table}

\subsubsection{Baseline Methods}
We compare five representative SID-based generative recommendation baselines, focusing on their SID tokenization strategies. \textbf{RK-Means}~\cite{QARM2025} is a non-parametric residual K-Means tokenizer that directly quantizes item embeddings into hierarchical SID sequences. \textbf{RQ-VAE}~\cite{TIGER2023}\footnote{``RQ-VAE'' refers to the unmodified RQ-VAE tokenizer from TIGER.} uses an RQ-VAE tokenizer~\cite{RQVAE2022} to derive hierarchical Semantic IDs and trains a T5-based~\cite{T52020} generative recommender to predict them autoregressively. \textbf{LETTER}~\cite{LETTER2024} extends the RQ-VAE tokenizer with collaborative alignment to incorporate interaction information and with code-diversity regularization to mitigate code-assignment imbalance. \textbf{QuaSID}~\cite{QuaSID2026} introduces collision-aware tokenizer training by distinguishing different collision types and applying severity-aware separation constraints. \textbf{MQL4GRec}~\cite{MQL4GRec2025} builds a quantitative language representation and resolves collisions through greedy distance-based reallocation, providing a zero-collision baseline. For a fair comparison, all tokenizers use the same textual item embeddings and produce length-$L{=}4$ SIDs over a codebook size of $V{=}256$.

\subsubsection{Evaluation Metrics}
In common practice, Hit@$K$ and NDCG@$K$ ($K \in {5, 10}$) are used as the standard evaluation protocol for generative recommendation. Previous work in recommendation systems evaluation has repeatedly emphasized that protocol choices can distort reported results~\cite{SampledMetrics2020,Reproducibility2019,BaselineDifficulty2019}. However, under SID collision, these metrics are evaluated at the SID level and can overestimate true item-level recommendation quality. We therefore use the collision-corrected ItemHit@$K$ and ItemNDCG@$K$ defined in Section~\ref{sec:cce} as the primary metrics throughout this paper. Table~\ref{tab:collision_inflation} quantifies the inflation of standard Hit@10 relative to ItemHit@10 across four datasets and four native tokenizers.

\subsubsection{Implementation Details}
The entire experimental pipeline is built on the official repository \footnote{\url{https://github.com/HonghuiBao2000/LETTER}} of LETTER, which provides both the T5-based generative recommender (training and inference) and reference implementations of the RQ-VAE and LETTER tokenizers. For a fair comparison across tokenizer families, the collaborative signal used by LETTER's collaborative regularization is the same PPMI+SVD embedding (Section~\ref{sec:cf}) shared by all other baselines. The RK-Means tokenizer is adapted from the residual K-Means implementation of OpenOneRec \footnote{\url{https://github.com/Kuaishou-OneRec/OpenOneRec}}. For QuaSID, whose official implementation has not been released, we re-implement the method strictly following the paper~\cite{QuaSID2026}. For MQL4GRec, we use the implementation of its tokenizer in the official repository \footnote{\url{https://github.com/zhaijianyang/MQL4GRec}}.

All experiments use a unified T5-based~\cite{T52020,Transformer2017} generator trained from scratch, so the comparison is controlled by the tokenizer rather than the autoregressive model. We use Qwen3-Embedding-8B~\cite{Qwen3Emb2025} to extract a textual embedding $\mathbf{x}_i \in \mathbb{R}^{d_t}$ ($d_t{=}4096$) for each item $i$ from its title and description. All tokenizers use $L = 4$ levels with codebook size $V = 256$. RK-Means runs 20 K-means iterations per level. RQ-VAE-family variants (TIGER, LETTER, QuaSID) train an MLP encoder with codeword dimension $32$ for up to $20\text{K}$ epochs using Adam (learning rate $10^{-3}$, batch size $8192$, commitment weight $\beta = 0.25$, K-means initialization, patience-$5$ early stopping with evaluation every $2\text{K}$ epochs). ZCR is applied post-quantization to all +ZCR variants. The T5-based generator has $4$ encoder/decoder layers with hidden dimension $d_{\text{model}}{=}128$, feed-forward dimension $1024$, $6$ attention heads, and dropout $0.1$. At inference, we use a beam search of width $20$. We repeat the main experiment with three random seeds ${42, 123, 2026}$, and report the mean performance in Table~\ref{tab:main_results}. All experiments are conducted on a single NVIDIA RTX PRO 6000 Blackwell GPU.

  \begin{table*}[t]
    \centering
    \caption{Native vs.\ +ZCR performance under collision-corrected item-level metrics. Red subscripts denote relative changes over the paired native tokenizer. MQL4GRec is a zero-collision reference without a native counterpart. Results are averaged over three seeds. Per-cell standard deviations across seeds range from $0.0001$ to $0.0032$ (median $0.0011$)}
    \label{tab:main_results}
    \begingroup
    \setlength{\tabcolsep}{3.5pt}
    \renewcommand{\arraystretch}{1.02}
  \begin{tabular*}{\textwidth}{@{\extracolsep{\fill}}llcccccccc@{}}
  \toprule
  \multirow{2}{*}{\textbf{Tokenizer}} & \multirow{2}{*}{\textbf{Variant}}
  & \multicolumn{2}{c}{\textbf{Scientific}}
  & \multicolumn{2}{c}{\textbf{Cell}}
  & \multicolumn{2}{c}{\textbf{Beauty}}
  & \multicolumn{2}{c}{\textbf{Yelp}} \\
  \cmidrule(lr){3-4}\cmidrule(lr){5-6}\cmidrule(lr){7-8}\cmidrule(lr){9-10}
  & & \shortstack{\textbf{ItemHit}\\\textbf{@5}} & \shortstack{\textbf{ItemHit}\\\textbf{@10}}
    & \shortstack{\textbf{ItemHit}\\\textbf{@5}} & \shortstack{\textbf{ItemHit}\\\textbf{@10}}
    & \shortstack{\textbf{ItemHit}\\\textbf{@5}} & \shortstack{\textbf{ItemHit}\\\textbf{@10}}
    & \shortstack{\textbf{ItemHit}\\\textbf{@5}} & \shortstack{\textbf{ItemHit}\\\textbf{@10}} \\
  \midrule
  \textbf{RK-Means} & native
    & 0.0421 & 0.0654 & 0.0383 & 0.0524 & 0.0335 & 0.0547 & 0.0191 & 0.0313 \\
  \textbf{RK-Means} & +ZCR
    & 0.0579\rup{37.4} & 0.0812\rup{24.1}
    & 0.0516\rup{34.6} & 0.0694\rup{32.4}
    & 0.0403\rup{20.5} & 0.0662\rup{21.1}
    & 0.0217\rup{13.7} & 0.0347\rup{11.1} \\
  \textbf{RQ-VAE} & native
    & 0.0468 & 0.0665 & 0.0446 & 0.0574 & 0.0365 & 0.0576 & 0.0215 & 0.0340 \\
  \textbf{RQ-VAE} & +ZCR
    & 0.0585\rup{25.0} & 0.0800\rup{20.4}
    & 0.0497\rup{11.6} & 0.0657\rup{14.4}
    & 0.0386\rup{5.6} & 0.0608\rup{5.6}
    & 0.0230\rup{7.0} & 0.0361\rup{6.0} \\
  \textbf{LETTER} & native
    & 0.0524 & 0.0767 & 0.0474 & 0.0628 & 0.0372 & 0.0610 & 0.0240 & 0.0405 \\
  \textbf{LETTER} & +ZCR
    & 0.0571\rup{9.0} & 0.0791\rup{3.1}
    & 0.0496\rup{4.6} & 0.0660\rup{5.0}
    & 0.0389\rup{4.5} & 0.0625\rup{2.5}
    & 0.0263\rup{9.5} & 0.0442\rup{9.0} \\
  \textbf{QuaSID} & native
    & 0.0544 & 0.0749 & 0.0490 & 0.0656 & 0.0392 & 0.0644 & 0.0221 & 0.0369 \\
  \textbf{QuaSID} & +ZCR
    & 0.0567\rup{4.1} & 0.0789\rup{5.3}
    & 0.0496\rup{1.4} & 0.0663\rup{1.0}
    & 0.0402\rup{2.5} & 0.0645\rup{0.2}
    & 0.0240\rup{8.6} & 0.0381\rup{3.2} \\
  \textbf{MQL4GRec} & built-in
    & 0.0426 & 0.0606 & 0.0453 & 0.0596 & 0.0361 & 0.0570 & 0.0219 & 0.0353 \\
  \midrule
  \multirow{2}{*}{\textbf{Tokenizer}} & \multirow{2}{*}{\textbf{Variant}}
  & \multicolumn{2}{c}{\textbf{Scientific}}
  & \multicolumn{2}{c}{\textbf{Cell}}
  & \multicolumn{2}{c}{\textbf{Beauty}}
  & \multicolumn{2}{c}{\textbf{Yelp}} \\
  \cmidrule(lr){3-4}\cmidrule(lr){5-6}\cmidrule(lr){7-8}\cmidrule(lr){9-10}
  & & \shortstack{\textbf{ItemNDCG}\\\textbf{@5}} & \shortstack{\textbf{ItemNDCG}\\\textbf{@10}}
    & \shortstack{\textbf{ItemNDCG}\\\textbf{@5}} & \shortstack{\textbf{ItemNDCG}\\\textbf{@10}}
    & \shortstack{\textbf{ItemNDCG}\\\textbf{@5}} & \shortstack{\textbf{ItemNDCG}\\\textbf{@10}}
    & \shortstack{\textbf{ItemNDCG}\\\textbf{@5}} & \shortstack{\textbf{ItemNDCG}\\\textbf{@10}} \\
  \midrule
  \textbf{RK-Means} & native
    & 0.0282 & 0.0357 & 0.0277 & 0.0322 & 0.0213 & 0.0281 & 0.0123 & 0.0163 \\
  \textbf{RK-Means} & +ZCR
    & 0.0429\rup{52.0} & 0.0504\rup{41.0}
    & 0.0400\rup{44.4} & 0.0457\rup{41.9}
    & 0.0255\rup{19.7} & 0.0338\rup{20.3}
    & 0.0141\rup{14.5} & 0.0183\rup{12.7} \\
  \textbf{RQ-VAE} & native
    & 0.0315 & 0.0379 & 0.0336 & 0.0378 & 0.0239 & 0.0306 & 0.0141 & 0.0181 \\
  \textbf{RQ-VAE} & +ZCR
    & 0.0445\rup{40.9} & 0.0514\rup{35.6}
    & 0.0391\rup{16.2} & 0.0442\rup{17.0}
    & 0.0252\rup{5.7} & 0.0324\rup{5.6}
    & 0.0148\rup{5.4} & 0.0190\rup{5.2} \\
  \textbf{LETTER} & native
    & 0.0361 & 0.0439 & 0.0355 & 0.0405 & 0.0239 & 0.0315 & 0.0155 & 0.0208 \\
  \textbf{LETTER} & +ZCR
    & 0.0443\rup{22.9} & 0.0514\rup{17.1}
    & 0.0388\rup{9.1} & 0.0440\rup{8.8}
    & 0.0256\rup{7.1} & 0.0331\rup{5.2}
    & 0.0168\rup{8.2} & 0.0225\rup{8.3} \\
  \textbf{QuaSID} & native
    & 0.0391 & 0.0457 & 0.0367 & 0.0421 & 0.0254 & 0.0334 & 0.0142 & 0.0189 \\
  \textbf{QuaSID} & +ZCR
    & 0.0438\rup{12.1} & 0.0509\rup{11.5}
    & 0.0388\rup{5.5} & 0.0441\rup{4.8}
    & 0.0263\rup{3.7} & 0.0341\rup{2.0}
    & 0.0154\rup{8.2} & 0.0199\rup{5.1} \\
  \textbf{MQL4GRec} & built-in
    & 0.0324 & 0.0382 & 0.0359 & 0.0405 & 0.0235 & 0.0302 & 0.0141 & 0.0184 \\
    \bottomrule
    \end{tabular*}
    \endgroup
  \end{table*}

\subsection{Effect of ZCR on Corrected Item-Level Performance (RQ3)}
\label{sec:main_results}
We next evaluate whether zero-collision reassignment preserves or improves corrected item-level performance of each tokenizer. For each tokenizer and dataset, we train the generative recommender under the same training protocol on the native SID assignment and on its ZCR version, and report the CCE item-level metrics (ItemHit@$K$ and ItemNDCG@$K$).

Table~\ref{tab:main_results} reports the corrected item-level performance. Each ZCR row is paired with the corresponding native row, with the red subscript indicating the relative change. MQL4GRec serves only as a built-in zero-collision reference because it is already collision-free by design.

The largest improvement appear for RK-Means and RQ-VAE, where native collision rates are higher. For example, RK-Means on Scientific improves IH@10 by 24.1\%. These larger improvements are consistent with the CCE scoring mechanism because native collision-group hits receive fractional item-level credit, whereas ZCR makes matched target SIDs unique and removes this discount. The relatively modest improvements of LETTER and QuaSID can be attributed to their lower native collision rates, which leave fewer last-level SIDs to be modified by ZCR and thereby limit the potential performance improvements. As shown in Table~\ref{tab:collision_inflation}, the collision rates range from $1.07\%$ to $8.59\%$ for LETTER and from $0.59\%$ to $6.23\%$ for QuaSID.

\subsection{Bias in SID-Level Tokenizer Evaluation (RQ1 and RQ2)}
\label{sec:eval_bias}

After evaluating item-level performance under zero-collision reassignment, we now examine why conventional SID-level evaluation can distort tokenizer comparisons. This section focuses on native tokenizer outputs before collision removal, where shared SID sequences cause Hit@$K$ and NDCG@$K$ to over-credit predictions that match collision groups rather than individual items. We first quantify the resulting metric inflation (RQ1), and then examine whether correcting this inflation changes the relative ranking of SID tokenizers (RQ2).

\subsubsection{Collision-Induced Metric Inflation (RQ1)}
\label{sec:collision_inflation}
\begin{table}[t]
  \centering
  \caption{Comparison of four existing tokenizers in terms of collision rate (Coll.\%), recommendation performance (H@10, IH@10), and inflation ratio (Infl.\%) before collision-resolution post-processing.}
  \label{tab:collision_inflation}
  \resizebox{\columnwidth}{!}{
  \begin{tabular}{llcccc}
  \toprule
  \textbf{Dataset} & \textbf{Tokenizer} & \textbf{Coll.\%} & \textbf{H@10} &
  \textbf{IH@10} & \textbf{Infl.\%} \\
  \midrule
  \multirow{4}{*}{Scientific}
      & RK-Means & 30.52 & 0.1330 & 0.0654 & 103.36 \\
      & RQ-VAE   & 20.42 & 0.1147 & 0.0665 & 72.48 \\
      & LETTER   &  1.24 & 0.0804 & 0.0767 & 4.82 \\
      & QuaSID   &  5.41 & 0.0768 & 0.0749 & 2.54 \\
    \midrule
    \multirow{4}{*}{Cell}
      & RK-Means & 28.52 & 0.0830 & 0.0524 & 58.40 \\
      & RQ-VAE   & 14.04 & 0.0702 & 0.0574 & 22.30 \\
      & LETTER   &  3.39 & 0.0673 & 0.0628 & 7.17 \\
      & QuaSID   &  3.25 & 0.0684 & 0.0656 & 4.27 \\
    \midrule
    \multirow{4}{*}{Beauty}
      & RK-Means & 21.58 & 0.0818 & 0.0547 & 49.54 \\
      & RQ-VAE   &  8.52 & 0.0693 & 0.0576 & 20.31 \\
      & LETTER   &  1.07 & 0.0664 & 0.0610 & 8.85 \\
      & QuaSID   &  0.59 & 0.0663 & 0.0644 & 2.95 \\
    \midrule
    \multirow{4}{*}{Yelp}
      & RK-Means & 15.73 & 0.0415 & 0.0313 & 32.59 \\
      & RQ-VAE   &  9.05 & 0.0393 & 0.0340 & 15.59 \\
      & LETTER   &  8.59 & 0.0457 & 0.0405 & 12.84 \\
      & QuaSID   &  6.23 & 0.0400 & 0.0369 & 8.40 \\
  \bottomrule
  \end{tabular}
  }
\end{table}

Table~\ref{tab:collision_inflation} reports how much conventional SID-level Hit@10 overestimates ItemHit@10 on the same native tokenizer outputs. Inflation is larger for higher-collision tokenizers, reaching its maximum for RK-Means on Scientific: Hit@10 is 0.1330, compared with an ItemHit@10 of 0.0654, a relative inflation of 103.36\%. Even lower-collision tokenizers such as LETTER and QuaSID exhibit measurable inflation on some datasets, confirming that collision bias is not specific to any single tokenizer family.

\subsubsection{Tokenizer Ranking Changes under Item-Level Evaluation (RQ2)}
\label{sec:ranking_changes}

The inflation in Table~\ref{tab:collision_inflation} is large enough to change tokenizer rankings when the same tokenizers are ordered by H@10 and IH@10. On three datasets (Scientific, Cell, and Beauty), RK-Means leads under SID-level H@10 but drops to last under corrected IH@10; after correction, LETTER ranks first on Scientific, while QuaSID ranks first on Cell and Beauty. The same pattern appears in pairwise comparisons: RQ-VAE ranks above LETTER by H@10 on Scientific, Cell, and Beauty, but LETTER ranks above RQ-VAE by IH@10 on the same datasets; on Yelp, LETTER already ranks above RQ-VAE under H@10. These reversals show that conventional SID-level metrics can yield misleading tokenizer rankings when collision rates differ, so faithful tokenizer comparison requires item-level correction rather than SID-level matching.

\subsection{Minimum-Cost Reassignment for Zero-Collision SIDs (RQ3)}

For the cost comparison in Table~\ref{tab:zcr}, the ``greedy'' column applies the nearest-codeword reassignment of MQL4GRec~\cite{MQL4GRec2025}. Within each prefix group, greedy first groups items with identical native last-level code values $s_i^{(L)}$. For each such group, it keeps the item with the smallest native assignment distance $D[i, s_i^{(L)}]$, and reassigns the remaining items sequentially to their nearest unused last-level codes within the same prefix group. This local sequential rule is contrasted with ZCR, which jointly minimizes the group-level reassignment cost in Eq.~\ref{eq:cr_cost}.

\begin{table}[t]
    \centering
    \caption{Reassignment cost under the zero-collision constraint. $n_{\text{reass}}$ denotes the number of reassigned items, while $\sum \Delta D$ denotes the total increase in assignment cost.}
    \label{tab:zcr}
    \resizebox{\columnwidth}{!}{
    \begin{tabular}{llcrrr}
    \toprule
    \multirow{2}{*}{Tokenizer} & \multirow{2}{*}{Dataset} & \multirow{2}{*}{$n_{\text{reass}}$} & \multicolumn{2}{c}{$\sum \Delta D$} & \multirow{2}{*}{Rel. drop} \\
    \cmidrule(lr){4-5}
    & & & greedy & ZCR & \\
    \midrule
    \multirow{4}{*}{RK-Means}
    & Scientific &    875  &  38{,}864 & \textbf{ 35{,}674} & $8.21\%$  \\
    & Cell       & 8{,}519 & 276{,}488 & \textbf{252{,}486} & $8.68\%$  \\
    & Beauty     & 1{,}574 &  63{,}409 & \textbf{ 55{,}678} & $12.19\%$ \\
    & Yelp       & 2{,}029 &  41{,}689 & \textbf{ 38{,}074} & $8.67\%$  \\
    \midrule
    \multirow{4}{*}{RQ-VAE}
    & Scientific &    535  &  9.64 & \textbf{ 8.39} & $12.95\%$ \\
    & Cell       & 4{,}017 & 324.64 & \textbf{279.06} & $14.04\%$ \\
    & Beauty     &    827  & 71.39 & \textbf{57.97} & $18.80\%$ \\
    & Yelp       & 2{,}574 & 46.45 & \textbf{40.22} & $13.40\%$ \\
    \bottomrule
    \end{tabular}}
\end{table}

Table~\ref{tab:zcr} shows that ZCR reduces reassignment cost on every reported tokenizer--dataset pair. Relative to greedy reassignment, ZCR reduces cost by 8.21\%--12.19\% for RK-Means and 12.95\%--18.80\% for RQ-VAE. These reductions show that the minimum-cost objective in Eq.~\ref{eq:cr_cost} reduces the distortion of reassignment relative to greedy reassignment while enforcing the same zero-collision constraint. MQL4GRec is omitted as a tokenizer row in this cost analysis because it is already collision-free by design.

\subsubsection{Case Study of Minimum-Cost Reassignment in Collision Group}
\label{sec:zcr_case_study}

\begin{figure}[t]
  \centering
  \includegraphics[width=0.9\columnwidth]{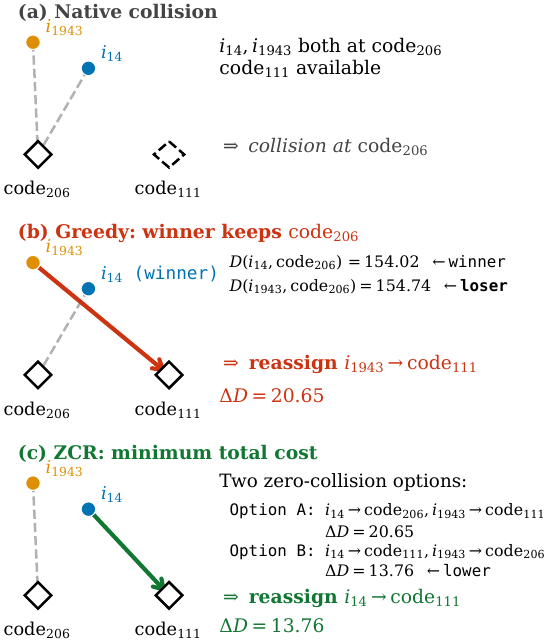}
  \caption{Case study of reassignment within one Beauty/RK-Means collision group (prefix $[127,244,179]$). (a) Native collision before reassignment. (b) Greedy sequential reassignment. (c) ZCR minimum-cost reassignment, whose joint minimum is identified by Eq.~\ref{eq:cr_cost} over all $V$ last-level codes. Positions are schematic, and costs are computed in the original residual space.}
  \Description{Three vertically stacked schematics compare native collision, greedy reassignment, and minimum-cost reassignment for items 14 and 1943 and codes 206 and 111. In panel (a), both items are assigned to code 206, creating a collision, while code 111 is unused. Panel (b) resolves the collision greedily by keeping the item closer to code 206 and reassigning item 1943 to code 111. Panel (c) instead compares the total cost of the two collision-free assignments and reassigns item 14 to code 111 because this yields the smaller increase in assignment cost.}
  \label{fig:zcr_case_study}
\end{figure}

We illustrate the reassignment mechanism on one real collision group from the RK-Means tokenizer for the Beauty dataset. The prefix $[127,244,179]$ contains items $i_{14}$ and $i_{1943}$, both natively assigned to the last-level code $s_{14}^{(L)} = s_{1943}^{(L)} = 206$. Resolving this collision requires one item to receive a different last-level code, while the other item can keep code $206$.

Greedy first compares the native assignment distances to code $206$, with $D[14,206] = 154.02$ and $D[1943,206] = 154.74$. Because item $i_{14}$ is closer to the occupied code, greedy keeps item $i_{14}$ at code $206$ and reassigns item $i_{1943}$ to its unused last-level code (code $111$), giving $\Delta D = 20.65$ (Figure~\ref{fig:zcr_case_study}(b)). ZCR instead solves Eq.~\ref{eq:cr_cost} over the same prefix group, under the requirement that items in the group receive distinct last-level codes. The minimum-cost assignment still changes only one item, but it keeps item $1943$ at code $206$ and assigns code $111$ to item $14$, giving $\Delta D = 13.76$ (Figure~\ref{fig:zcr_case_study}(c)). This case highlights the mechanism behind the aggregate reductions in Table~\ref{tab:zcr}. Greedy decides which item keeps the occupied code using only a local distance comparison, whereas ZCR minimizes the assignment cost over the prefix group. Consistent with this case, Table~\ref{tab:zcr} reports a 12.19\% reduction from greedy to ZCR.

\subsection{Further Analysis}
\label{sec:further_analysis}

\subsubsection{Collaborative Fusion on Zero-Collision RK-Means}
\label{sec:cf}

\begin{figure}[t]
  \centering
  \includegraphics[width=\columnwidth]{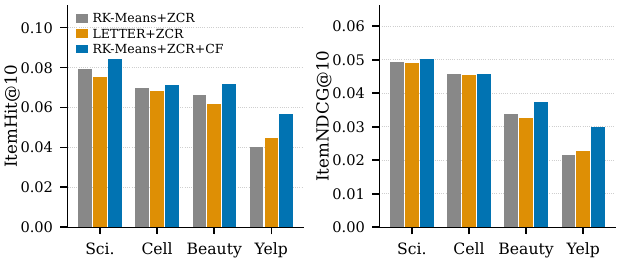}
  \caption{Effect of collaborative signal under zero-collision evaluation.}
  \Description{Two grouped bar charts compare three zero-collision tokenizer variants on Scientific, Cell, Beauty, and Yelp. The left chart reports ItemHit at 10 and the right reports ItemNDCG at 10. Each dataset compares RK-Means plus ZCR, LETTER plus ZCR, and RK-Means plus ZCR plus collaborative fusion. The collaborative-fusion variant achieves the highest ItemHit at 10 on all four datasets, with the largest gain over RK-Means plus ZCR on Yelp. ItemNDCG differences are small on Scientific and Cell, while the collaborative-fusion variant achieves higher ItemNDCG on Beauty and Yelp.}
  \label{fig:cf}
\end{figure}

Among the baselines we compare, LETTER incorporates collaborative signals into its RQ-VAE tokenizer through collaborative alignment regularization, suggesting that interaction-derived information, in addition to textual embeddings, can improve tokenization quality. We examine a simpler fusion before quantization, and compare it with both a textual-only RK-Means+ZCR and LETTER+ZCR under the same zero-collision evaluation. Figure~\ref{fig:cf} contrasts three variants: RK-Means+ZCR, which applies RK-Means to textual embeddings, followed by ZCR; LETTER+ZCR, which applies LETTER to textual embeddings (with its built-in collaborative alignment regularization), followed by ZCR; and RK-Means+ZCR+CF, which applies RK-Means to the fused textual and collaborative representation, followed by ZCR. All three variants use the same generator and are evaluated with zero-collision SID assignments.

\paragraph{PPMI-SVD Collaborative Embedding.}
Following~\cite{PPMI-SVD2014}, we obtain a $256$-dim collaborative item embedding $\mathbf{x}_i^{\mathrm{cf}}$ by truncated SVD on the PPMI matrix of item co-occurrences from user training sequences (window $3$, last $2$ items per user held out to prevent train-test leakage), with L2-normalization. To construct the fused tokenizer input $\mathbf{z}_i \in \mathbb{R}^{d_t}$, we also L2-normalize the textual embedding $\mathbf{x}_i$, form the concatenation $[\mathbf{x}_i;\, \alpha\, \mathbf{x}_i^{\mathrm{cf}}]$ (fusion hyperparameter $\alpha = 0.5$), and apply PCA, keeping the top $d_t$ principal components.

Adding collaborative fusion to RK-Means+ZCR improves IH@10 on all four datasets, with gains from $+2.2\%$ to $+63.1\%$. RK-Means + ZCR+CF also achieves the highest IH@10 among the three variants in every dataset, outperforming LETTER+ZCR by $+4.6\%$ to $+26.2\%$. The largest gain appears on Yelp, where item text is limited to business names and categories rather than the richer product metadata available in the Amazon datasets. This pattern is consistent with collaborative signals contributing more where textual metadata is less discriminative. On Yelp, RK-Means+ZCR underperforms LETTER+ZCR ($0.0347$ vs.\ $0.0442$), showing that collaborative information is useful beyond textual RK-Means alone. RK-Means+ZCR+CF further improves IH@10 to $0.0566$.

\begin{figure*}[t]
  \centering
  \includegraphics[width=0.8\textwidth]{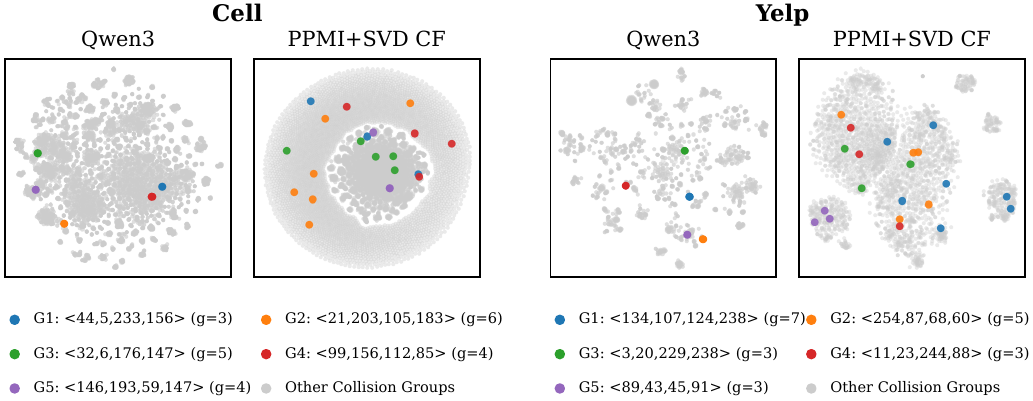}
  \caption{$t$-SNE visualization of item embeddings on Cell and Yelp, under Qwen3 textual embeddings and PPMI+SVD collaborative embeddings. Different colors of data points indicate distinct highlighted SID groups, and light grey points belong to other collision groups.}
  \Description{Four scatterplots compare two-dimensional projections of textual and collaborative item embeddings for Cell and Yelp. For each dataset, the left panel shows Qwen3 textual embeddings and the right panel shows collaborative embeddings derived from positive pointwise mutual information and singular value decomposition. Light grey points denote items from other collision groups. Items within each highlighted collision group appear tightly clustered in the textual projection but more dispersed in the collaborative projection. The legends identify each group by its Semantic ID sequence and group size.}
  \label{fig:collision_viz}
\end{figure*}

\subsubsection{Visualization of Collision Groups between Textual and Collaborative Spaces}

To investigate the origin of SID collisions illustrated by the example of similar 3D printers in Figure~\ref{fig:sid_collision}, we employ t-SNE~\cite{tSNE2008} to project items into two representation spaces: the textual space (Qwen3 embedding) and the collaborative space (PPMI+SVD collaborative embedding in Section~\ref{sec:cf}), as shown in Figure~\ref{fig:collision_viz}. Specifically, for two datasets (Cell and Yelp), we color-highlight the five collision groups (size $g{\ge}3$) with the smallest mean intra-group Qwen3 cosine distance per dataset. The highlighted groups appear as near-overlapping clusters in the textual space, consistent with the text-only RK-Means tokenizer assigning them the same SID. In contrast, the same items are more separated in the collaborative space, suggesting that collaborative signals contain item-distinguishing information not captured by the pure-semantic SID assignment. Thus, the visualization illustrates a source of SID collision: textually similar items may be merged by semantic quantization even when their interaction patterns remain distinguishable.

\section{Related Work}

\subsection{SID-based Generative Recommendation}

SID-based generative recommendation maps each item to a sequence of discrete semantic IDs and trains an autoregressive model to generate the SID sequence of the target item from user interactions~\cite{TIGER2023, P52022}. In contrast to sequential recommenders that predict items through learned item ID embeddings~\cite{GRU4Rec2016,SASRec2018,BERT4Rec2019,zhang2025semantic,zhang2022rethinking}, this paradigm transfers semantic structure from pretrained item embeddings into the SID space, so that similar items share codeword prefixes and the model can decode SIDs hierarchically~\cite{CoLa2026}.

Existing item tokenization methods can be broadly categorized into \emph{foundational quantization methods} and their \emph{recsys-aware augmentations}. For foundational quantization, TIGER~\cite{TIGER2023} introduces residual-quantized variational autoencoders (RQ-VAE)~\cite{RQVAE2022,VQVAE2017} to map item embeddings into discrete SID sequences. QARM~\cite{QARM2025,OpenOneRec2025,GRID2025} replaces the variational autoencoder with deterministic residual $k$-means. VQ-Rec~\cite{VQRec2023} adopts optimized product quantization (OPQ)~\cite{OPQ2014} for transferable cross-domain SID representations. Based on these foundations, some methods inject recsys-aware signals to improve the quality of SIDs. LETTER~\cite{LETTER2024} augments RQ-VAE with semantic, collaborative, and diversity regularizers. EAGER~\cite{EAGER2024} introduces a two-stream architecture that contrastively aligns behavior and semantic streams. LMIndexer~\cite{LMIndexer2024} learns the tokenizer through self-supervised language modeling. Beyond the tokenizer design itself, ETEGRec~\cite{ETEGRec2025} jointly optimizes the tokenizer and the generative recommender. Recent studies have advanced SID tokenization along multiple complementary directions: aligning collaborative and semantic signals at the token level~\cite{TokenRec2025,TCA4Rec2026,FACE2025}, constructing personalized or context-aware SIDs~\cite{Pctx2025,IDs2Sem2026}, decomposing item semantics into coarse-to-fine levels~\cite{CoFiRec2025}, and developing cascaded sparse–dense or continuous-token representations~\cite{SparseDense2025,DiffuGR2026}. However, their evaluations still rely on SID-level matching, which is valid only when SID assignments are zero-collision.

\subsection{SID Collision and Existing Mitigation}

SID collision occurs when multiple items are assigned the same SID sequence. Many SID tokenizers (e.g., RQ-VAE~\cite{TIGER2023}, OPQ~\cite{RPG2025}) for generative recommendation typically build on vector quantization methods, which allow similar vectors to share the same discrete code. However, the recommendation operates at the item level: textually similar items often correspond to distinct interaction patterns, and VQ's many-to-one mapping conflates them into the same SID sequence.

Existing approaches to mitigating SID collision follow two routes: methods that modify tokenizer training and methods that resolve collisions after tokenization. The first route modifies the tokenizer loss or architecture to reduce collisions during tokenizer training. These include uniform semantic mapping for non-conflicting index assignment~\cite{LCRec2024}, qualification-aware separation for different collision types~\cite{QuaSID2026}, uniqueness losses combined with hierarchical disentanglement~\cite{HiDVAE2025}, and adaptive overlap regulation that separates only harmful collisions~\cite{AdaSID2026}. The second route leaves the trained tokenizer unchanged and resolves collisions afterward. These include appending an extra disambiguating code to each SID sequence~\cite{TIGER2023}, greedy distance-based reallocation~\cite{MQL4GRec2025}, and relaxed nearest-centroid algorithms that produce unique IDs without appended codes~\cite{PSI2025}. 

These methods reduce or eliminate collisions during or after tokenization, but collisions can remain in native tokenizer outputs, and collision rates vary substantially across tokenizers~\cite{SIDInspector2026}. Faithful evaluation across tokenizers under different collision rates has received little attention~\cite{RegressionAccumulation2026}. CCE fills this gap by computing item-level metrics directly from generated SID sequences, without requiring either retraining or modifying the tokenizer. Among methods that resolve collisions after tokenization, appending extra non-semantic codes increases the SID length~\cite{TIGER2023}, while greedy reallocation introduces excessive changes to the original SID assignment~\cite{MQL4GRec2025}. ZCR instead keeps the SID length unchanged and resolves collisions by performing minimum-cost reassignment within each prefix group, yielding a uniform setting for cross-tokenizer comparisons.

\section{Limitations}
Our evaluation has three limitations. First, when multiple items share an SID, CCE assigns them a uniform credit because the generated SID cannot distinguish them. A popularity- or context-aware assignment could refine this, but this work is a faithful evaluation across tokenizers rather than within-group ranking. Second, ZCR changes only the last code and leaves the prefix fixed, so its optimality depends on each prefix group fitting in the size of the codebook, $\max_{\mathbf{p}} |\mathcal{P}(\mathbf{p})| \le V$ (Eq.~\ref{eq:codebook_capacity}). This holds in all our experiments. When this condition fails (a small codebook or many near-duplicate items under a single prefix), enlarging $V$ or incorporating multimodal item features is the most direct solution. It is noted that reassigning across multiple levels would also work, but it changes the prefix structure that ZCR is meant to preserve. Third, we conduct experiments on public Amazon and Yelp datasets using one generator architecture. This keeps the tokenizer the only moving part, but the inflation size and ranking flips may differ on larger industrial datasets.

\section{Conclusion}
We present a faithful evaluation framework that comprises collision-aware metrics (CCE) and a zero-collision reassignment method (ZCR). CCE computes item-level credit from generated SID sequences without retraining, correcting conventional SID-level matching under SID collision. ZCR resolves collisions through minimum-cost reassignment, providing a uniform zero-collision setting for tokenizer comparison. Across four datasets and five representative SID tokenizers, collision rates reach 30.52\%, and SID-level Hit@10 inflation reaches 103.36\%. Item-level re-evaluation reverses several tokenizer rankings implied by SID-level metrics. These results show that predicting the target SID is not always equivalent to identifying the target item and that faithful tokenizer evaluation requires item-level correction or explicit zero-collision assignment. In future work, we will explore tokenizer designs that guarantee item-level identifiability while preserving semantic structure, and examine whether SID collision persists in other SID-based generative recommendation paradigms.

\appendix

\section{Empirical Verification of the Last-Level Codebook Capacity Condition}
\label{sec:capacity-appendix}

Section~\ref{sec:optimality} establishes that the last-level reassignment of ZCR achieves the global minimum over assignments preserving the first $L{-}1$ codewords and modifying only the last level, with zero collision guaranteed when the capacity condition $\max_\mathbf{p}|\mathcal{P}(\mathbf{p})| \le V$ (Eq.~\ref{eq:codebook_capacity}) holds. Table~\ref{tab:prefix_group_max} reports the maximum and mean prefix-group sizes on native SID assignments for the four tokenizers across all four datasets. MQL4GRec is omitted because its native output is already zero-collision, and we do not apply ZCR to it.

\begin{table}[h]
\centering
\caption{Maximum and mean prefix-group sizes $|\mathcal{P}(\mathbf{p})|$ per tokenizer and datasets on native SID assignments, reported as max (mean). All maximum values are below the codebook size $V=256$.}
\label{tab:prefix_group_max}
\begin{tabular}{lrrrr}
    \toprule
    Tokenizer & Scientific & Cell & Beauty & Yelp \\
    \midrule
    RK-Means & 29 (1.39) & 91 (1.52) & 27 (1.38) & 65 (1.29) \\
    RQ-VAE   & 23 (1.30) & 73 (1.21) & 18 (1.14) & 65 (1.12) \\
    LETTER   &  7 (1.02) & 18 (1.05) & 24 (1.04) & 65 (1.27) \\
    QuaSID   &  8 (1.10) & 14 (1.03) &  4 (1.02) & 65 (1.05) \\
    \bottomrule
\end{tabular}
\end{table}

Two observations follow from these results. First, the maximum prefix-group size in the Yelp column is identically $65$ across all four tokenizers. Inspection of this prefix group reveals $65$ items with identical title (``Starbucks'') and description (``Food, Coffee \& Tea'') located at different addresses. A second prefix group of $56$ items exhibits the same invariance across all rows. The Yelp maxima therefore arise from intrinsic metadata duplication in the dataset, which explains why the same maximum appears across tokenizers. Second, the mean prefix-group sizes range from $1.02$ to $1.52$, indicating that most prefix groups contain only one item and that the maxima are driven by a small number of large prefix groups. The capacity condition $\max_\mathbf{p}|\mathcal{P}(\mathbf{p})| \le V = 256$ is therefore satisfied with substantial margin across all four tokenizers and four datasets, and the last-level-only reassignment of ZCR achieves zero-collision SID assignments in every case, consistent with the analysis in Section~\ref{sec:optimality}.


\section*{GenAI Usage Disclosure}
We used a generative language model (GPT-5.5) to assist with grammar correction and sentence-level polishing. All core contributions, including all ideas, method design, experimental setup, and analysis, were made by the authors.

\bibliographystyle{ACM-Reference-Format}
\bibliography{main-base}

\end{document}